\documentclass[pre,twocolumn,showpacs,preprintnumbers,amsmath,amssymb]{revtex4}
\usepackage{graphicx,gensymb}
\usepackage{dcolumn}
\usepackage{bm}

\begin{document}

\title{Condensation of elastic energy in two-dimensional packing of wire}
\author{C. C. Donato, and M. A. F. Gomes}
\affiliation{Departamento de F\'{\i}sica, Universidade Federal de
Pernambuco, 50670-901, Recife - PE - Brazil}

%\date{\today}

\begin{abstract}
Forced packing of a long metallic wire injected into a
two-dimensional cavity leads to crushed structures involving a
hierarchical cascade of loops with varying curvature radii. We study
the distribution of elastic energy stored in such systems from
experiments, and high-resolution digital techniques. It is found
that the set where the elastic energy of curvature is concentrated
has dimension $D_\mathcal{S} = 1.0 \pm 0.1$, while the set where the
mass is distributed, has dimension $D =1.9 \pm 0.1$.
\end{abstract}


\maketitle

\section{\label{sec1} INTRODUCTION}

\indent Packing problems are of noteworthy importance to many
branches of industry and science. Theoretical, experimental, and
technological investigations of these problems have attracted much
attention in connection with number theory, coding and group
theory, analog-to-digital conversion and data compression,
$d$-dimensional crystallography and condensed-matter physics in
general, as well as in dual theory and superstring
theory~\cite{1}. Frequently we are interested in the strategy to
obtain the most efficient way to pack a large number of equal
extended units, say spheres, in $d$-dimensional Euclidean space.
Even in the area of two dimensional circle packing, there is a
myriad of open important problems of theoretical and applied
interest~\cite{2}. A less studied problem in two-dimensional
packing, which has great technological interest, is the search of
a rigid system with a low packing fraction. An example of this
last class of system is a single layer structure of loops obtained
when a crushed wire is confined in a two-dimensional
cavity~\cite{3}. This specific case yields rigid metallic
structures with a maximum average occupation fraction of $14\%$ of
the area irrespective of the packing~\cite{4}. In fact, this
maximum packing fraction cannot vary appreciably with the
material, because it is controlled by the approximately universal
ratio of the shear modulus to the modulus of elasticity of the
wire~\cite{5}.

On the other hand, many physical systems present the phenomenon of
condensation of energy, that is, a $D$-dimensional system in a space
of dimension $d \geq D$ can have some type of energy concentrated in
a small fraction of its total volume. In principle, the subset
$\mathcal{S}$ where this particular energy is localized can have a
fractal dimension $D_\mathcal{S} \leq D$. At this point it is
opportune to introduce the concept of support. In mathematics, the
support of a quantity is the set $\mathcal{S}$ where that quantity
is nonzero. Many soft-condensed matter systems as proteins~\cite{7},
polymers in general~\cite{8}, and diffusion limited
aggregates~\cite{9} have in physical space fractal structures with
dimension $D < 3$. They are formed from large number of units glued
together by short-range covalent chemical forces and/or van der
Waals forces. The existence of these systems themselves is a proof
of condensation of chemical and van der Waals energies on low
dimensional supports. In these cases, the distribution of
short-range forces follows the respective structure of the systems,
and as a consequence $D_\mathcal{S}$ is equal to $D$.

Condensation of elastic energy, in particular, has raised recently
a growing interest in connection with crumpled sheets, and
$m$-dimensional elastic manifolds~\cite{10,11}. Other problems
related to the physics of crumpled structures with the topology of
a sheet have been investigated in the last two decades, as
examples one can mention studies on basic statistical
aspects~\cite{12,13,14,15,16}, on stress and strain relaxation in
crumpled thin sheets~\cite{17}, among others~\cite{18}. On the
contrary, the physics of crushed structures with one-dimensional
topology as exemplified by a squeezed ball of wire, has been much
less studied. Geometrical, statistical, and physical aspects of
crushed wires in three-dimensional space were previously examined
experimentally from the point of view of robust scaling laws.
Fractal dimensions associated with these disordered systems were
reported~\cite{19}.

In this article we study the condensation of elastic energy in
configurations of a copper wire injected into a planar
two-dimensional cavity, using experiments, scaling and
high-resolution digital techniques for data acquisition. The
structures studied here are remarkably different from crumpled
sheets or crushed wires in three dimensions.

\section{\label{sec2} EXPERIMENTAL DETAILS}

\begin{figure}[h!]
\resizebox{5.5cm}{!}{\includegraphics{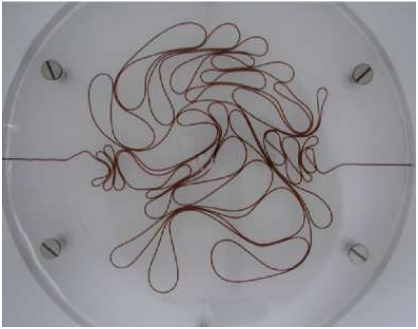}}
\caption{\label{fig1} A real packing configuration of $\#$19AWG
copper wire within the cavity of 20.0~cm of diameter.}
\end{figure}

Our experiments of packing of wire were performed in a
two-dimensional transparent cell consisting of the superposition
of two discs of plexiglass with a total height of 1.8~cm, and an
external radius of 15~cm. The internal circular cavity of the cell
had a radius $R_0 =$ 10.0~cm, and 0.11~cm of height, and it can
only accommodate configurations of a single layer of the folded
structure of the wire. The cavity of the cell was polished to
reduce the friction between wire and cavity. The $\#$19AWG copper
wire used in the experiments had a diameter $\zeta =$ 0.10~cm and
a varnished surface, in order to reduce wire-wire friction. Cavity
and wire operated in dry regime, free of any lubricant, and the
injection of wire into the cell was made through two channels
along the diameter as shown in Figure~\ref{fig1}. The photographs
were taken with a digital camera with a resolution of 3.2~Mpixel.
The digital images were transferred to a personal computer for
processing. This stage consists of digital filtering, data
validation, and conversion into binary images.

Each packing experiment begins fitting a straight wire in the
channels and subsequently pushing manually and uniformly the wire
on both sides of the cell toward the interior of the cavity with a
velocity of order of 1~cm/s. However, the observed phenomena are
widely independent of the injection speed for the whole interval
of injection velocity compatible with a manual process. The
packing of an elastic wire in a cavity involves a hierarchical
cascade of loops heterogeneously distributed in space as
exemplified in Figure~\ref{fig1}. In this figure the packing
fraction is $p=0.14$, corresponding to a length $L =$~440~cm of
wire. This fraction $p$ is defined as the ratio of the projected
area of the crushed wire to the area of the cavity and is equal to
$\zeta L/\pi R_0^2$. When the length of wire within the cavity
increases in the interval $p > 0.10$, the difficulty in the
injection begins to rise, with a corresponding reduction in the
velocity of injection. For $p$ near the maximum packing fraction,
$p_{max}$, the difficulty in the injection rises abruptly and the
crushed structures finally become rigid: for $p > p_{max}$, the
crushed wire becomes completely jammed within the cavity and it is
impossible to continue the injection of wire. The particular
moment when the injection velocity goes rapidly to zero leads to a
tight-packing (TP) configuration for the crushed wire
(Figure~\ref{fig1}). The average maximum packing fraction in the
experiment is $p_{max}=0.14 \pm 0.02$, corresponding to an aspect
ratio $L/\zeta$ = $(\pi R_0^2 p/\zeta^2) = (4.40 \pm 0.63) \times
10^3$, irrespective the existence of lubrication in the cavity.
The reader can observe that the sharp creases and ridges found in
crumpling of sheets are absent in the two-dimensional packing of
wire shown in Figure~\ref{fig1}. In order to better visualize the
packing process considered here, we refer to our previous
work~\cite{3,4}.

The packing of a copper wire in our experiments is not perfectly
reversible~\cite{3,4}, however the degree of irreversibility is
small, and considerable elastic energy remains stored in the
cavity, as we confirm from the observation of the strong uncoiling
of the wire when the cell is disconnected after a very long
period. As we can see from Figure~\ref{fig1}, the loops are units
that have a bulge in one extremity, and in the other extremity the
two branches of the loop merge. The bulge is formed by different
contiguous small arcs characterized by different radii of
curvature $\rho$, where the elastic energy of curvature,
proportional to $(1/\rho)^2$~\cite{11}, can be stored in different
degrees. The extremity of each loop opposite to the bulge is
associated with arcs of very large curvature radii and, as a
consequence, its capacity to store elastic energy is comparatively
very small. In this article we concentrate in this particular type
of elastic energy, since the energy cost for stretching the wire
is much larger than the bending cost.

\begin{figure}[h!]
\resizebox{7cm}{!}{\includegraphics{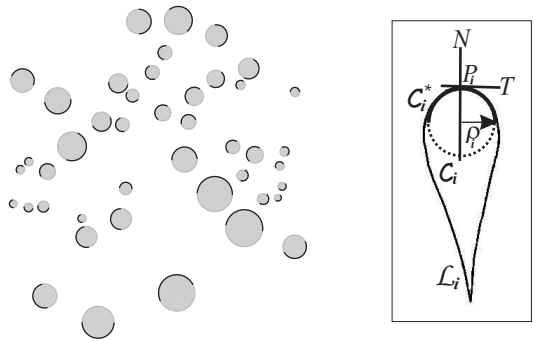}}

\caption{\label{fig2} Support $\mathcal{S}$ of condensation of
elastic energy for the packing in Figure~\ref{fig1}. Arcs are
regions of the folded structure of wire that contribute to
$\mathcal{S}$. Shading denotes osculatory circles. Inset: loop
$\mathcal{L}_i$ with the quantities defined in the text.}
\end{figure}

Now we define, in the context of this work, the support
$\mathcal{S}$ where the elastic energy of the crushed wire is
distributed (Figure~\ref{fig2}, inset): for each loop
$\mathcal{L}_i$, we found the point $P_i \in \mathcal{L}_i$ where
the loop presents the smaller radius of curvature, $\rho_i$. Due
to the geometry of the loops, this point $P_i$ is localized in the
bulge of the loop. The tangent $T$ and its corresponding normal
$N$ at $P_i$ are found, and the associated osculatory circle
$\mathcal{C}_i$ is inserted, as indicated in Figure~\ref{fig2}. If
$\rho_i$ is smaller than an arbitrary threshold $\rho_c$, the arc
$\mathcal{C}_i^*$ defined by the half part of $\mathcal{C}_i$ (one
quarter of the circumference from each side of the normal, with
the mid-point of $\mathcal{C}_i^*$ at $P_i$) is considered an
element of $\mathcal{S}$. If, on the contrary, $\rho_i > \rho_c$
we consider that $\mathcal{L}_i$ does not contribute to the
support $\mathcal{S}$. Then, $\mathcal{S}$ is a collection
$\{\mathcal{C}_i^*\}$ of semi circumferences, as shown in
Figure~\ref{fig2}. This figure refers to the support of energy
associated with the crushed wire shown in Figure~\ref{fig1}, and
the threshold adopted in this case was $\rho_c = 22 \zeta
=$~2.2~cm.

\section{\label{sec3} RESULTS AND DISCUSSION}

\begin{figure}[h!]
\hspace{-0.5cm} \resizebox{8.5cm}{!}{\includegraphics{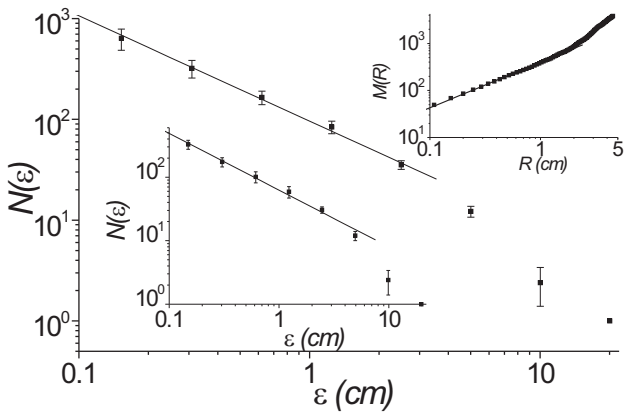}}
\caption{\label{fig3} Main plot: number $N(\epsilon)$ of boxes of
size $\epsilon$ needed to cover the energy support $\mathcal{S}$
as a function of $\epsilon$. Upper inset: the corresponding
mass-length plot. Lower inset: $N(\epsilon)$ when the arcs of
$\mathcal{S}$ have angular aperture of $90\degree$ instead of
$180\degree$. All plots are averages on five experiments.}
\end{figure}

For testing scale invariance in the support $\mathcal{S}$, we
apply both the box counting and the average mass-length
methods~\cite{20}. Box counting was performed by overlapping a
grid of size $\epsilon$  over the set and counting the number
$N(\epsilon)$ of squares of size $\epsilon$ needed to cover
$\mathcal{S}$. For fractal objects $N(\epsilon) \sim
\epsilon^{-D_\mathcal{S}}$, where $D_\mathcal{S}$ is the fractal
dimension. The mass-length relation measures the dependence
between the mass $M(R)$ of $\mathcal{S}$ within circles of radii
$R$ fully included in the cavity after averaging on many centers
and all configurations. In our case, $M(R)$ is given by the number
of pixels within the circles. For a scale-free set, it is expected
a scaling relation of the type $M(R) \sim R^{D_\mathcal{S}'}$,
where $D_\mathcal{S}'$ is the mass exponent. The main plot in
Figure~\ref{fig3} illustrates the results obtained with the
box-covering method; in this case we get $N(\epsilon) \sim
\epsilon^{-D_\mathcal{S}}$, with $D_\mathcal{S} = 1.0 \pm 0.1$
(average on the 5 experimental configurations). On the other hand,
from the power law scaling for $M(R)$ shown in the upper inset of
the figure we find $D_\mathcal{S}' = 1.0 \pm 0.1$ in agreement
with the box counting exponent $D_\mathcal{S}$. These exponents
however, are markedly different from the mass exponent $D = 1.9
\pm 0.1$, found for the entire mass distribution of wire within
the cavity in the situation of maximum packing
fraction~\cite{3,4}. These tests suggest that the support where
energy is condensed, according to our previous definition, has a
low dimension, which seems to be a well-defined property of the
system. The values of $D_\mathcal{S}$ and $D_\mathcal{S}'$ are,
within the error bars, independent of both $\rho_c$ and the
fraction $\phi$  of the perimeter of each osculatory circle
$\mathcal{C}_i$ included in the support $\mathcal{S}$ (in
Figure~\ref{fig2}, we have $\phi =0.5$, corresponding to an
angular aperture of $180\degree$). To illustrate this last aspect,
we give in the lower inset of Figure~\ref{fig3} the plot of
$N(\epsilon)$ for the energetic arcs when the angular aperture of
the arcs is reduced to $90\degree$ ($\phi = 0.25$); in this case,
$N(\epsilon) \sim \epsilon^{-D_\mathcal{S}}$, with $D_\mathcal{S}
=0.9 \pm 0.1$.

\begin{figure}[h!]
\hspace{-1.25cm} {\includegraphics{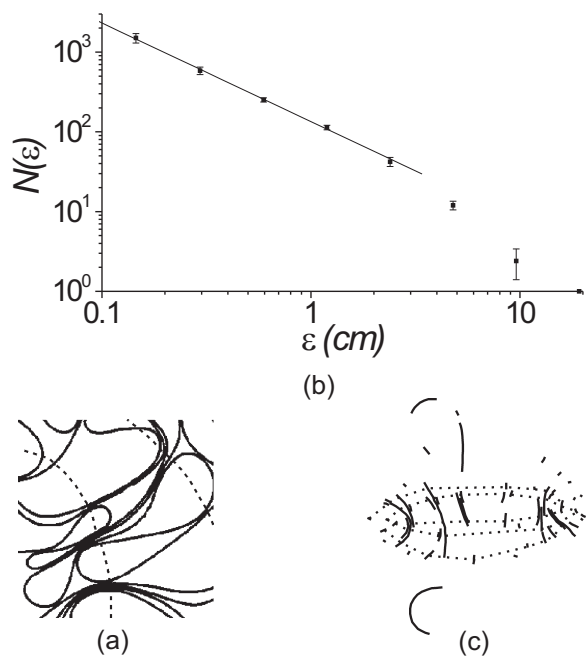}}
\caption{\label{fig4} (a) Detail of loops and wire-wire contacts.
(b) Box counting function $N(\epsilon)$ for wire-wire contacts for
the ensemble of crushed wires. (c) Typical full pattern of
wire-wire contacts. Dashed lines connecting wire-wire contacts are
possibly paths for propagation of forces across the structure.}
\end{figure}

Finally we present an analysis of how the wire-wire contacts are
geometrically distributed across the cavity. In the packing of
wires, as we can observe from Figure~\ref{fig1}, there are
tangency regions, i.e. regions presenting $n$-fold wire-wire
contacts. In an $n$-fold contact, $n + 1$ pieces of wire come
together into direct contact as detailed in Figure~\ref{fig4}(a).
Each contact region has an area given by the length of the
tangency region multiplied by the width $(n+1) \zeta$ of the
corresponding number $n+1$ of segments of wire contributing to the
contact.The plot in Figure~\ref{fig4}(b) shows the result of a box
counting analysis for the patterns of wire-wire contacts for the
ensemble of configurations of crushed wire examined in this
article. The experimental data are well described by a power law
decay suggesting a fractal dimension $D_{ww} = 1.2 \pm 0.1$, a
value close to $D_\mathcal{S}$ and $D_{\mathcal{S}'}$. We
speculate that the wire-wire contacts define arcs of propagating
forces across the structure as suggested by the five dashed lines
in the lower inset of Figure~\ref{fig4}(c). The impossibility to
eliminate completely the friction along wire-wire interfaces leads
to dashed lines only approximately orthogonal to the contact
regions. The numerical value obtained for $D_{ww}$ seems in
agreement with the idea that forces are transmitted primarily
along one-dimensional structures in noncrystalline systems, as
e.g. observed in granular materials~\cite{21,22,23}.

\section{\label{sec4} CONCLUSION}

In summary, we have studied experimentally in detail the geometry
of the support where the elastic energy of curvature is stored in
a packing of crushed wire in a two-dimensional cavity. Robust
scaling laws connecting variables of interest are reported, and
the associated critical exponents are determined. Based on the
experimental data for two-dimensional systems, it is found that
the elastic energy is concentrated on a set of dimensionality
close to unity. It is possible that in a $d$-dimensional cavity,
the elastic energy could be condensed equally on a low-dimensional
support with dimension $D_{\mathcal{S}} \cong 1$. Elastic
materials as steel wires and nylon fishing lines exhibit patterns
of folds as that shown in Figure~\ref{fig1}. This indicates that
$D_{\mathcal{S}}$ in these cases has the same value reported in
this article.

\section*{Acknowledgement}
This work was supported in part by Conselho Nacional de
Desenvolvimento Cient\'{\i}fico e Tecnol\'ogico, and Programa de
N\'ucleos de Excel\^encia (Brazilian Agencies). M. A. F. G.
acknowledges the interest and stimulus from H. J. Herrmann, and J.
S. Andrade Jr. We are also grateful to M. Engelsberg for a
critical reading of this manuscript.

\end{document}